\documentstyle[epsf]{article}%
  \def\baselinestretch{1.8}
\begin{document}
\title{Distances and classification of amino acids for
different protein secondary structures}
\author{Xin Liu${^\dag}$,Li-mei Zhang${^\ast}$, Shan Guan${^\ddag}$,
Wei-Mou Zheng${^\dag}$\\
${^\dag}${\it Institute of Theoretical Physics, China,
Beijing 100080, China}\\
${^\ast}${\it School of Science at North Jiaotong University,
Beijing 100044, China}\\
${^\ddag}${\it Center of Bioinformations at Peking University,
Beijing 100871, China}}
\date{}
\maketitle

\begin{abstract}
Window profiles of amino acids in protein sequences are taken as a
description of the amino acid environment. The relative entropy or
Kullback-Leibler distance derived from profiles is used as a
measure of dissimilarity for comparison of amino acids and
secondary structure conformations. Distance matrices of amino acid
pairs at different conformations are obtained, which display a
non-negligible dependence of amino acid similarity on
conformations. Based on the conformation specific distances
clustering analysis for amino acids is conducted.

\end{abstract}

\leftline{PACS number(s): 87.10.+e,02.50.-r}%
\bigskip

\section{Introduction}
The similarity of amino acids($aa$) is the basis of protein
sequence alignment, protein design and protein structure
prediction. Several scoring schemes have been proposed based on
amino acid similarity. The mutation data matrices of Dayhoff
\cite{dhf} and the substitution matrices of Henikoff \cite{hh} are
standard choices of scores for sequence alignment and amino acid
similarity evaluation. However, these matrices, focusing on the
whole protein database, pay little attention on protein secondary
structures($ss$). How the amino acid similarity is influenced by
different secondary structures is an interesting question.
Furthermore, understanding the differences can help us in protein
sequence analysis.

Despite efforts in uncovering the information encoded in the
primary structure, we still cannot read the language describing
the final 3D fold of an active biological macromolecule. Compared
with the DNA sequence, a protein sequence is generally much
shorter, but the size of the alphabet is five times larger. A
proper coarse-graining of the 20 amino acids into fewer clusters
for different conformation is important for improving the
signal-to-noise ratio when extracting information by statistical
means.

It is our purpose to propose a simple scheme to study amino acid
similarity from amino acid string statistics. Information about
the environment for an amino acid at a certain conformation state
may be provided by statistics of residue strings or windows
centered at the amino acid. The success of window-based approaches
such as GOR \cite{gor} for secondary structure prediction
validates the use of such statistics. We shall derive a measure
for the difference of amino acid pairs based on the distance of
probability distributions, and investigate how the difference is
dependent on conformations.

\section{Amino acid distances}

Our discussion will be heavily based on the distance between two
probability distributions. A well defined measure of the distance
is the Kullback-Leibler (KL) distance or relative entropy
\cite{kl,kl1,kl2}, which, for two distributions $\{p_i\}$ and
$\{q_i\}$, is given by
\begin{equation}
d(\{p_i\}, \{q_i\})= \sum_i p_i \log(p_i/q_i).
\end{equation}
It corresponds a likelihood ratio, and, if $p_i$ is expanded around
$q_i$, its leading term is the $\chi^2$ distance:
\begin{equation}
d_\chi (\{p_i\}, \{q_i\})= \sum_i (p_i -q_i)^2/p_i.
\label{chi}
\end{equation}
It is often to use the following symmetrized form for the KL distance
\begin{equation}
D(\{p_i\}, \{q_i\})= \hbox{$\frac 12$}[d(\{p_i\}, \{q_i\})+
d(\{q_i\}, \{p_i\})].
\label{kl}
\end{equation}

The distributions to be considered here come from window
statistics. For a given amino acid residue $a_i=x$ at the
conformation state $\alpha$ in a sequence $a_1a_2\cdots
a_i\cdots$, we take the string $a_{-n+i}a_{-n+i+1}\cdots a_i\cdots
a_{i+n}$ of width $(2n+1)$ as a window. Denote by $N_k(y|x,\alpha
)$ the count of residue $y$ at the $k$-th site from the center of
such windows. As in GOR, only the conformation of the central
residue is concerned. A quantity derived from $N_k(y|x,\alpha)$ is
\begin{equation}
N(x,\alpha )=\sum_y N_k(y|x,\alpha ),
\end {equation}
which, as the total count of residue $x$ at the conformation $\alpha$,
is independent of $k$. The conditional probability distribution
$P_k(y|x,\alpha )$ is estimated as
\begin{equation}
P_k(y|x,\alpha )=\frac {N_k(y|x,\alpha )} {N(x,\alpha )}.
\end {equation}
The weight matrix ${\bf M}_{20\times 2n}$ with its entries being
$P_k(y|x,\alpha )$ is the so-called residue profile of $x$ at
$\alpha$. Such profiles are used in window-based approaches, e.g.
GOR and artificial neural network algorithm \cite{lapa}.

We expect that on average the correlation between the central residue
and an outer site decays when they become far apart in sequence. To
examine the correlation, we consider a large window width of 21, i.e.
$n=10$, and take the `noise' background to be the following
average:
\begin{equation}
Q(y|x,\alpha )=\hbox{$\frac 16$}\left[\sum_{k=-10}^{-8}P_k(y|x,\alpha )
+ \sum_{k=8}^{10}P_k(y|x,\alpha )\right].
\end {equation}
The KL distance $D_{k;x,\alpha}(\{P_k(y|x,\alpha )\},
\{Q(y|x,\alpha )\})$ provides a measure of
the correlation between the central site and site $k$. As we shall see,
for our purpose of amino acid comparison a narrow window of a strong
correlation with width of 7 is used to describe amino acid enviroment.

Using distribution $P_k(y|x,\alpha )$ from window statistics to
characterize amino acid residues, we define the distance of residue
pair $x$ and $y$ at the same conformation $\alpha$ as the following
sum of KL distances
\begin{equation}
D_{xy;\alpha}= \sum_{k=\pm 1,\pm 2, \pm 3} D(\{P_k(z|x,\alpha )\},
\{P_k(z|y,\alpha )\}). \label{dxy}
\end{equation}
Similarly, to explore the difference of the same residue $x$ at
different conformations $\alpha$ and $\beta$, we may define the
distance
\begin{equation}
D_{\alpha\beta ;x}= \sum_{k=\pm 1,\pm 2, \pm 3} D(\{P_k(z|x,\alpha
)\}, \{P_k(z|x,\beta )\}). \label{dab}
\end{equation}

By means of the residue pair distances we can further study the
classification of amino acids. With the KL distance, we may
define the cluster distance in a way consistent with that for
residue pairs. For example, we characterize the cluster consisting
of residues $x$ and $y$ by the `coarse-grained' probability
\begin{equation}
P_k(z|x\&y,\alpha )=\frac {N_k(z|x,\alpha )+N_k(z|y,\alpha )}
{N(x,\alpha )+N(y,\alpha )}.
\label{x+y}
\end{equation}
We then may define the distance between this cluster and some
other residues or clusters. With cluster distance defined, the
cluster analysis can be used to reduce amino acid alphabets.

\section{Results}

Our analysis is performed on a data set taken from the database
PDB\_SELECT\cite{pslt,psltseq} of nonredundant protein sequences
with known structures. The sequences share amino acid identity
less than 25\%. We keep only the non-membrane sequences with their
lengths between 80 and 420. The secondary structure assignment is
taken from the DSSP database \cite{dssp}. As in GOR, we use the
following reduction of the 8 DSSP states to 4 states of helix(h),
sheet(e), coil(c) and turn(t): $H, G, I\to h$, $E\to e$, $X, S,
B\to c$ and $T\to t$. The counts of each amino acid at the reduced
four different conformation states are given in Table 1.

We first estimate probability distributions of residues for each
central residue at a given conformation. At this step, the window
width is 21. We then calculate distances
$D_{k;x,\alpha}(\{P_k(y|x,\alpha )\}, \{Q(y|x,\alpha )\})$ of
these distributions to their corresponding noise distributions.
The results are shown in Figs.~1 to 4, each of which is for one
conformation of the central residue. The 20 curves in each figure
correspond to 20 central amino acids. Due to the sample size difference, curves
are not directly comparable. (Roughly speaking, under the null
hypothesis of identical distribution the $\chi^2$ distance should
be scaled with the sample size, so a small sample size would give
a relatively large distance.) However, a decay is clearly seen
when the site $k$ become far away from the center. For more
discussions on correlations we refer reader to
\cite{weiss,weiss1}. As seen from most curves of the figures,
distances at the 6 sites nearest to the center are significantly
larger than those at window border sites. We shall use window
width of 7 for further comparison of amino acids.

It is natural to expect that similar residues would have similar
window statistics. Thus, the KL distance between two residue
profiles provides a measure of their similarity, i.e. a small KL
distance implies a large similarity. We calculate the KL distance
matrices $D_{xy;\alpha}$ for residue pairs at different
conformations with formula (\ref{dxy}). The results are given in
Tables 2 and 3, where entries have been multiplied by a factor
200. With the distributions (\ref{x+y}) defined for clusters, we
further perform the simplest bottom-up approach of hierachical
clustering for residues, by starting from 20 clusters of single
residues, and then joining two nearest clusters step by step until
a single cluster is obtained. The results of clustering are given
in Tables 4 to 7. Since the dendritic trees returned from
clustering are less informative, for visualization we introduce
graphs where vertices are the 20 amino acids, and an edge exists
between a pair of amino acids if and only if their distance is
below some preset threshold. Graphs obtained from the distance
matrices are shown in Figs.~5 to 8, where vertices with no
connecting edges are neglected.

In sequence pair alignment we often do not have structure information
of both sequences. With the structure information ignored, we have the
mixed counts
\begin{equation}
N_k(y|x)=\sum_\alpha N_k(y|x,\alpha ),
\end{equation}
from which we calculate the residue pair distances averaged over
conformations. The distance matrix obtained is given in Table 8.
We have also calculated distances (\ref{dab}) to compare different
conformations. Distances between any two conformations for various
residues are listed in Table 9.

\section{Discussions}

Figures 1 to 4 illustrate the dependence of outer sites in a window
on the center. Although in the KL distance we sum up effects on
individual residues from the center, we still can see the tendency
that the center is generally more strongly correlated with the
C-terminal sites than N-terminal sites. Furthermore, we may divide
the 20 amino acids into two groups with M, I, L, V, F, Y and W in
one, and the remainders in the other. They roughly correspond to
hydrophobic and hydrophilic groups. It is seen that for the coil
and turn conformations a hydrophobic center exhibits a stronger
correlation with outer sites than a hydrophilic center, while for
the sheet conformation a hydrophilic center exhibits a stronger
correlation.

It is interesting to make a comparison between the distance matrices
obtained here with the commonly used BLOSUM62 similarity score matrix.
A small distance implies a large similarity score. There are many
evidences showing the consistency between the distances and scores.
For example, residue pairs VI, IL, VL and ST have positive BLOSUM
scores and at the same time small distances. The graphs in Figs.~5
to 8 contain two connected subgraphs: one consists of I, L, V, F,
Y, and the other consists of S, T. This is another evidence of the
consistency. Generally, the averaged distance matrix is closer to
BLOSUM62 than the conformation specific ones. However, there do exist
some remarkable differences. For example,
residue pairs GT, QA, FV with negative scores have rather small
distances in either the conformation helix, or sheet or coil, while
pairs YH and NH with positive scores have rather large distances in
the helix conformation. Moreover, YH has a large distance in all the
four conformations.

BLOSUM matrices are derived from conserved amino acid patterns
called blocks. It is expected that for most score entries we
should see the consistency in at least one conformation specific
distance matrix. For a given residue pair, if residue profiles of
an amino acid center are very dissimilar for different
conformations, after averaging over conformations the pair
distance would generally become smaller. In this case, BLOSUM
scores and conformation specific distance need not be consistent
since the former contains no structure information.

Our results show some strong dependence of residue behavior on
conformations. For example, the distances of pairs CD and SI in
helix are about twice higher than in sheet. There are many residue
pairs displaying strong dependence of distances on conformations.
Table 9 views the conformation dependence from conformation pair
comparison. Indeed, the table indicates that for any conformation
pairs there are certain residues which behave very differently in
the two conformations. However, generally speaking, coil and turn
are quite similar.

In comparison of physicochemical properties of amino acids, the
abundance of amino acids is not taken into consideration. This is
also the case for the above defined distances. Other statistical
variables including the effect of sample size may be introduced.
One candidate is the $\chi^2$ statistic for identical
distributions. The analysis using this new statistic is under
study.

We expect that algorithms using multiple conformation specific matrices
should work better in sequence alignment. The popular Needleman-Wunsch
algorithm can be modified to include putative conformation for each
residue. This will be discussed elsewhere.

\begin{quotation}
{This work was supported in part by the Special Funds for Major
National Basic Research Projects and the National Natural Science
Foundation of China.}
\end{quotation}

\def
\baselinestretch{1}

Table 1. Sample sizes of each amino acid residue in different
protein secondary structures.

\smallskip
\footnotesize
\begin{tabular}{crrrr}
\hline
 & h & e & c & t \\ \hline
C&       690&        732&        822&        224\\
S&      2841&       1764&       3538&       1179\\
T&      2350&       2288&       3112&        762\\
P&      1173&        624&       3648&       1302\\
A&      5950&       2019&       2651&       1122\\
G&      1795&       1633&       4328&       3090\\
N&      1904&        922&       2692&       1388\\
D&      2841&       1029&       3621&       1424\\
E&      4773&       1514&       2325&       1172\\
Q&      2757&       1008&       1532&        653\\
H&      1132&        794&       1148&        426\\
R&      3108&       1469&       1948&        771\\
K&      3861&       1579&       2645&       1187\\
M&      1390&        693&        679&        223\\
I&      3169&       3333&       1719&        368\\
L&      6262&       3307&       2952&        850\\
V&      3233&       4461&       2330&        487\\
F&      2225&       1948&       1545&        444\\
Y&      1806&       1773&       1303&        459\\
W&       827&        632&        536&        173\\
\hline
\end{tabular}

\newpage

Table 2. Amino acid distance matrices for helices (bottom-left) and
turns (top-right). Entries have been multiplied by a factor 200.

\footnotesize{ \noindent $\begin{array}{crrrrrrrrrrrrrrrrrrrr}
C&     &106&116&135&118&145&134&132&121&129&111&118&124&154&134&121&119&104&123&215\\
S& 64&     & 23& 52& 29& 59& 35& 33& 36& 37& 54& 26& 36& 78& 61& 38& 49& 45& 40&100\\
T& 63& 13&     & 61& 33& 74& 40& 35& 39& 46& 62& 33& 37& 92& 63& 40& 46& 45& 38& 93\\
P& 81& 48& 49&     & 44& 99& 71& 69& 54& 62& 82& 47& 55&106& 89& 71& 71& 62& 66&132\\
A& 45& 21& 17& 63&     & 64& 38& 39& 32& 36& 58& 29& 33& 63& 64& 34& 46& 48& 43& 98\\
G& 57& 15&20& 52& 25&     & 32& 39& 54& 55& 57& 47& 52& 81& 79& 61& 88& 75& 70&115\\
N& 82& 14& 22& 67& 33& 26&     & 18& 30& 31& 44& 29& 31& 72& 68& 38& 63& 53& 36& 96\\
D&101& 17& 26& 56& 39& 32& 16&     & 25& 34& 44& 30& 29& 77& 58& 36& 54& 49& 37& 91\\
E& 82& 20& 25& 56& 27& 36& 22& 14&     & 33& 53& 37& 23& 73& 65& 46& 51& 59& 43&106\\
Q& 70& 16&21& 60&19& 28& 17& 21& 14&     & 51& 32& 38& 79& 66& 51& 62& 62& 49&100\\
H& 55&23&24& 55& 26& 26&33& 35& 34& 28&     & 51& 58& 90& 79& 54& 81& 70&55&113\\
R& 69&21&22& 59&21& 30& 22& 28& 24& 13& 28&     & 30& 71& 69& 38& 54& 49& 48&101\\
K& 80& 21& 25& 67& 28& 38& 22& 27& 23& 19& 38& 13&     & 81& 63& 38& 49& 51& 47&102\\
M& 48& 57& 45& 85& 23& 56& 75& 82& 64& 51& 50& 50& 60&     & 93& 62& 85& 93& 78&141\\
I& 43& 81& 65&104& 35& 78&104&116& 88& 76& 66& 73& 79& 22&     & 47& 55& 55& 54&104\\
L& 35& 65& 52& 90& 26& 62& 83& 99& 73& 59& 53& 56& 67& 15& 10&     & 49& 36& 31& 85\\
V& 37& 59& 44& 81& 22& 55& 77& 90& 67& 53& 52& 51& 60& 16& 12& 09&     & 46& 58& 99\\
F& 34& 67& 53& 87& 30& 61& 90& 99& 79& 69& 54& 66& 75& 22& 17& 12&15&     & 48&100\\
Y& 44& 43& 35& 77&23& 47& 64& 71& 55& 47&34& 47& 54& 26& 29&21&22& 16&     & 90\\
W& 49& 61& 53& 82& 35& 64& 87& 92& 72& 58& 57& 60& 71& 31& 35& 27& 29& 25& 24&     \\
 & C  &  S & T  &  P & A  &   G&  N &  D &   E&  Q & H  & R  & K  & M  &I   &L   &V   & F  &Y    &W
\end{array}$
}
\\
\normalsize

 Table 3. Amino acid distance matrices for sheets
 (bottom-left) and coils (top-right). Entries have been multiplied
 by a factor 200.\\

\noindent\footnotesize{ $\begin{array}{crrrrrrrrrrrrrrrrrrrr}
C&     &43& 51& 70& 36& 47& 51& 66& 50& 51& 46& 48& 54& 55& 51& 44& 42& 47& 48& 66\\
S& 42&     &10& 24& 17& 17& 16& 21& 20& 14& 30& 15& 19& 32& 38& 27& 24& 30& 32& 45\\
T& 49& 15&     &28& 19& 24& 16& 21& 17& 16& 34& 17& 17& 26& 32& 24& 22& 28& 31& 46\\
P& 68& 42& 46&     &37& 28& 25& 22& 31& 28& 48& 28& 31& 52& 67& 59& 49& 62& 61& 71\\
A& 33& 20& 24& 42&     &16& 22& 29& 16& 16& 30& 17& 21& 23& 27& 17& 15& 24& 25& 41\\
G& 35& 29& 37& 62& 16&     &18& 23& 31& 28& 36& 17& 31& 31& 44& 17& 27& 31& 32& 44\\
N& 51& 23& 27& 46& 30& 37&     &14& 19& 19& 30& 19& 20& 34& 42& 34& 31& 31& 34& 54\\
D& 54& 24& 31& 46& 32& 42& 23&     &22& 23& 39& 23& 22& 46& 56& 48& 41& 46& 51& 65\\
E& 60& 21&19& 48& 32& 47& 26& 24&     &14& 32& 14& 11& 25& 30& 25& 17& 25& 24& 39\\
Q& 52& 20&17& 53& 28& 41& 29& 30& 22&     &32& 14& 14& 30& 33& 26& 21& 26& 25& 38\\
H& 50& 27& 26& 54& 28& 33&34& 33& 30& 28&     & 29& 36& 51& 44& 41& 39& 44& 40& 67\\
R& 46&21&20& 44&20& 33& 32& 31& 21& 23& 22&     & 16& 31& 33& 24& 21& 28& 29& 47\\
K& 62& 29&20& 52& 30& 47& 35& 34& 20& 23& 35& 24&     & 34& 34& 28& 21& 29& 26& 50\\
M& 38& 45& 44& 65&24& 33& 52& 62& 50& 46& 44& 38& 52&     & 28& 22& 22& 27& 30& 39\\
I& 32& 38& 36& 62&24& 35& 56& 57& 49& 41& 40& 36& 43& 23&     & 12& 15& 16& 20& 34\\
L& 27& 37& 34& 58&19& 29& 50& 55& 45& 41& 37& 32& 43& 20& 09&     & 10& 12& 17& 33\\
V& 31& 35& 32& 58& 19& 27& 51& 57& 46& 40& 36& 32& 38& 22& 09& 10&     & 14& 17& 29\\
F& 29& 45& 44& 71&25& 33& 62& 67& 59& 47& 49& 42& 56& 28& 14& 12&15&     & 18&33\\
Y& 32& 35& 32& 64&24& 33& 51& 54& 47& 34&33& 31& 42& 29&13&13&15& 14&     & 31\\
W& 46& 57& 58& 71& 47& 60& 69& 76& 62& 52& 54& 57& 66& 48& 39& 39& 38&33&37&     \\
 & C  &  S & T  &  P & A  &   G&  N &  D &   E&  Q & H  & R  & K  & M  &I   &L   &V   & F  &Y    &W
\end{array}$
}
\\
\normalsize
\newpage
Table 4. Clustering of amino acid alphabets for helices. The first
column indicates the number of amino acid groups.

{\baselineskip=0pt
\begin{verbatim}
19 A D E K Q R S T N G H C F I LV M Y W P
18 A D E K Q R S T N G H C F ILV  M Y W P
17 A D E K Q R S T N G H C FILV   M Y W P
16 A D E K Q R ST  N G H C FILV   M Y W P
15 A D E K QR  ST  N G H C FILV   M Y W P
14 A D E KQR   ST  N G H C FILV   M Y W P
13 A D E KQRST     N G H C FILV   M Y W P
12 A D E KQRSTN      G H C FILV   M Y W P
11 A D EKQRSTN       G H C FILV   M Y W P
10 A DEKQRSTN        G H C FILV   M Y W P
 9 A DEKQRSTN        G H C FILVM    Y W P
 8 ADEKQRSTN         G H C FILVM    Y W P
 7 ADEKQRSTN         G H C FILVMY     W P
 6 ADEKQRSTNG          H C FILVMY     W P
 5 ADEKQRSTNGH           C FILVMY     W P
 4 ADEKQRSTNGH           C FILVMYW      P
 3 ADEKQRSTNGH           CFILVMYW       P
 2 ADEKQRSTNGHCFILVMYW                  P
\end{verbatim}}

Table 5. Clustering of amino acid alphabets for sheets. The first
column indicates the number of amino acid groups.

{\baselineskip=0pt
\begin{verbatim}
19 A G F IL V Y M D E Q S T R K H N C W P
18 A G F ILV  Y M D E Q S T R K H N C W P
17 A G FILV   Y M D E Q S T R K H N C W P
16 A G FILVY    M D E Q S T R K H N C W P
15 A G FILVY    M D E Q ST  R K H N C W P
14 A G FILVY    M D E QST   R K H N C W P
13 A G FILVY    M D EQST    R K H N C W P
12 A G FILVY    M D EQSTR     K H N C W P
11 A G FILVY    M D EQSTRK      H N C W P
10 AG  FILVY    M D EQSTRK      H N C W P
 9 AGFILVY      M D EQSTRK      H N C W P
 8 AGFILVYM       D EQSTRK      H N C W P
 7 AGFILVYM       D EQSTRKH       N C W P
 6 AGFILVYM       D EQSTRKHN        C W P
 5 AGFILVYM       DEQSTRKHN         C W P
 4 AGFILVYMDEQSTRKHN                C W P
 3 AGFILVYMDEQSTRKHNC                 W P
 2 AGFILVYMDEQSTRKHNCW                  P
\end{verbatim}}

Table 6. Clustering of amino acid alphabets for coils. The first
column indicates the number of amino acid groups.

{\baselineskip=0pt
\begin{verbatim}
19 A E K Q R ST N G D F L V I Y M H P W C
18 A E K Q R ST N G D F LV  I Y M H P W C
17 A E K Q R ST N G D FLV   I Y M H P W C
16 A E K Q R ST N G D FLVI    Y M H P W C
15 A EK  Q R ST N G D FLVI    Y M H P W C
14 A EKQ   R ST N G D FLVI    Y M H P W C
13 A EKQR    ST N G D FLVI    Y M H P W C
12 A EKQRST     N G D FLVI    Y M H P W C
11 AEKQRST      N G D FLVI    Y M H P W C
10 AEKQRSTN       G D FLVI    Y M H P W C
 9 AEKQRSTNG        D FLVI    Y M H P W C
 8 AEKQRSTNG        D FLVIY     M H P W C
 7 AEKQRSTNGD         FLVIY     M H P W C
 6 AEKQRSTNGDFLVIY              M H P W C
 5 AEKQRSTNGDFLVIYM               H P W C
 4 AEKQRSTNGDFLVIYMH                P W C
 3 AEKQRSTNGDFLVIYMHP                 W C
 2 AEKQRSTNGDFLVIYMHPW                  C
\end{verbatim}}
\newpage
Table 7. Clustering of amino acid alphabets for turns. The first
column indicates the number of amino acid groups.

{\baselineskip=0pt
\begin{verbatim}
19 A DN E K S T R Q L Y F V H G I P M W C
18 A DN E K ST  R Q L Y F V H G I P M W C
17 A DN EK  ST  R Q L Y F V H G I P M W C
16 A DNEK   ST  R Q L Y F V H G I P M W C
15 A DNEKST     R Q L Y F V H G I P M W C
14 A DNEKSTR      Q L Y F V H G I P M W C
13 ADNEKSTR       Q L Y F V H G I P M W C
12 ADNEKSTRQ        L Y F V H G I P M W C
11 ADNEKSTRQL         Y F V H G I P M W C
10 ADNEKSTRQLY          F V H G I P M W C
 9 ADNEKSTRQLYF           V H G I P M W C
 8 ADNEKSTRQLYFV            H G I P M W C
 7 ADNEKSTRQLYFVH             G I P M W C
 6 ADNEKSTRQLYFVHG              I P M W C
 5 ADNEKSTRQLYFVHGI               P M W C
 4 ADNEKSTRQLYFVHGIP                M W C
 3 ADNEKSTRQLYFVHGIPM                 W C
 2 ADNEKSTRQLYFVHGIPMW                  C
\end{verbatim}}

Table 8. Amino acid distances ignoring conformation.\\

 \noindent \footnotesize$\begin{array}{crrrrrrrrrrrrrrrrrrrr}
C&     & & & & & & & & & & & & & & & & & & & \\
S& 21&     & & & & & & & & & & & & & & & & & & \\
T& 25& 5&     & & & & & & & & & & & & & & & & & \\
P& 25& 9& 11&     & & & & & & & & & & & & & & & & \\
A& 29& 12& 12& 16&     & & & & & & & & & & & & & & & \\
G& 21& 8& 11& 11& 11&     & & & & & & & & & & & & & & \\
N& 25& 7& 9& 13& 12& 8&     & & & & & & & & & & & & & \\
D& 32& 9& 9& 15& 10& 11& 6&     & & & & & & & & & & & & \\
E& 40& 18& 18& 21& 11& 18& 14& 9&     & & & & & & & & & & & \\
Q& 34& 12& 12& 18& 8& 14& 10& 9& 8&     & & & & & & & & & & \\
H& 21& 13& 14& 17& 18& 14& 12& 15& 23& 17&     & & & & & & & & & \\
R& 31& 11& 13& 16& 7& 13& 11& 10& 9& 5& 15&     & & & & & & & & \\
K& 35& 15& 14& 18& 12& 16& 10& 9& 8& 10& 22& 8&     & & & & & & & \\
M& 33& 19& 16& 20& 10& 17& 18& 18& 19& 16& 24& 15& 18&     & & & & & & \\
I& 25& 16& 13& 16& 12& 14& 16& 17& 20& 18& 19& 16& 15& 10&     & & & & & \\
L& 26& 16& 14& 17& 9& 14& 16& 17& 19& 15& 20& 14& 15& 8& 4&     & & & & \\
V& 24& 10& 9& 13& 8& 9& 11& 12& 15& 13& 17& 12& 12& 10& 6& 6&     & & & \\
F& 22& 13& 11& 16& 13& 11& 14& 16& 20& 18& 18& 16& 15& 12& 6& 6& 6&     & & \\
Y& 24& 9& 9& 13& 13& 10& 11& 14& 19& 15& 14& 15& 14& 13& 8& 9& 7& 5&     & \\
W& 32& 20& 19& 20& 21& 17& 22& 25& 29& 23& 24& 24& 27& 18& 14& 13& 13& 10& 12&     \\
 & C  &  S & T  &  P & A  &   G&  N &  D &   E&  Q & H  & R  & K  & M  &I   &L   &V   & F  &Y    &W
\end{array}$\\
\newpage
 Table 9. Conformation pair distances for each amino acid. Entries
have been multiplied by a factor 200. (h: Helix, e: Sheet, c:
coil,
t: Turn.)\\

\noindent\footnotesize{ $\begin{tabular}{ccccccc} \hline
 &he &hc &ht &ec &et &ct \\
\hline
C&133&185&163&127&197&139\\
S& 93&129&124& 93&148& 73\\
T& 98&120&131&103&175& 96\\
P&172&118&121& 89&233&116\\
A&112&148&127&122&149& 73\\
G& 79&101& 80& 91&107& 57\\
N&126&145&118&106&152& 76\\
D&149&137&149& 93&174& 81\\
E&159&152&138&109&192& 73\\
Q&130&157&133& 93&143& 93\\
H&100&150&110&117&152& 98\\
R&131&146&128& 91&144& 85\\
K&137&149&128& 93&155& 88\\
M&130&161&147&126&156&135\\
I&138&180&134&118&130&110\\
L&143&162&113&127&148& 98\\
V&114&151&151& 98&147&101\\
F&120&150&111&107&115& 88\\
Y& 95&147& 96&111&117& 80\\
W&120&181&201&123&173&111\\
\hline
\end{tabular}$}
\\

\begin{figure}
\centerline{\epsfxsize=12cm \epsfbox{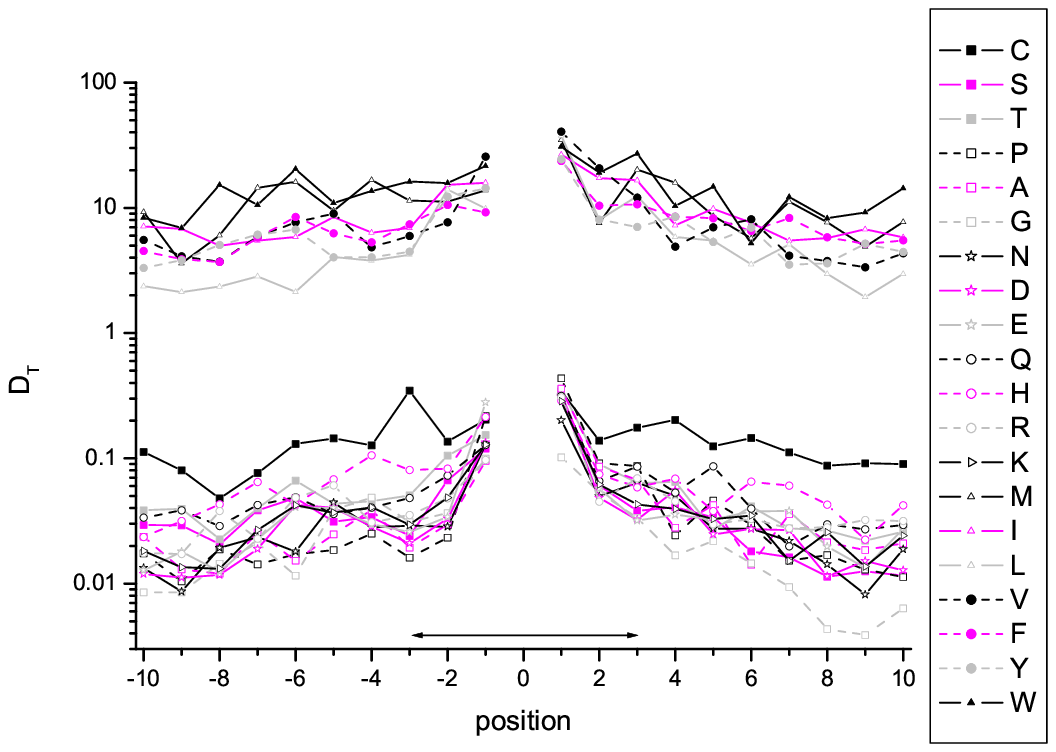}}
\caption{KL distances (doubled) of outer sites from their corresponding
noise background. Each curve is for an amino acid at the center labeled 0,
whose conformation is turn. For clarity, the curves for M,I,L,V,F,Y and
W have been shifted up by multiplying an extra factor 100. }
\label{fig1}
\end{figure}
\begin{figure}
\centerline{\epsfxsize=12cm \epsfbox{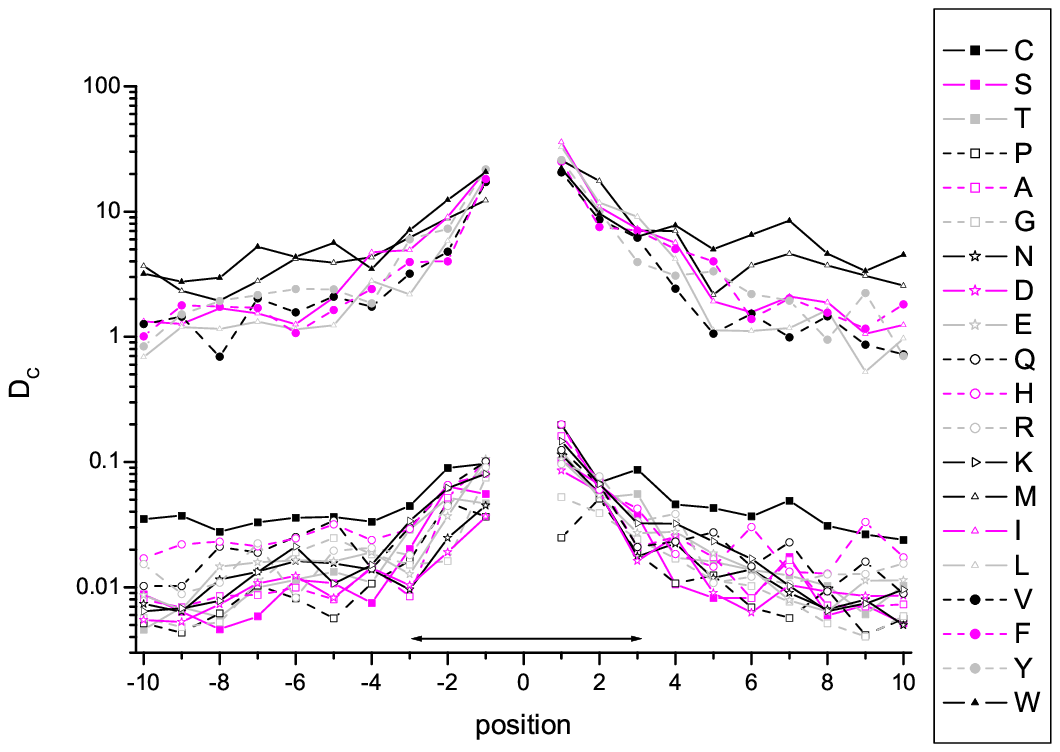}}
\caption{KL distances (doubled) of outer sites from their corresponding
noise background. Each curve is for an amino acid at the center labeled 0,
whose conformation is coil. For clarity, the curves for M,I,L,V,F,Y and
W have been shifted up by multiplying an extra factor 100.} \label{fig2}
\end{figure}
\begin{figure}
\centerline{\epsfxsize=12cm \epsfbox{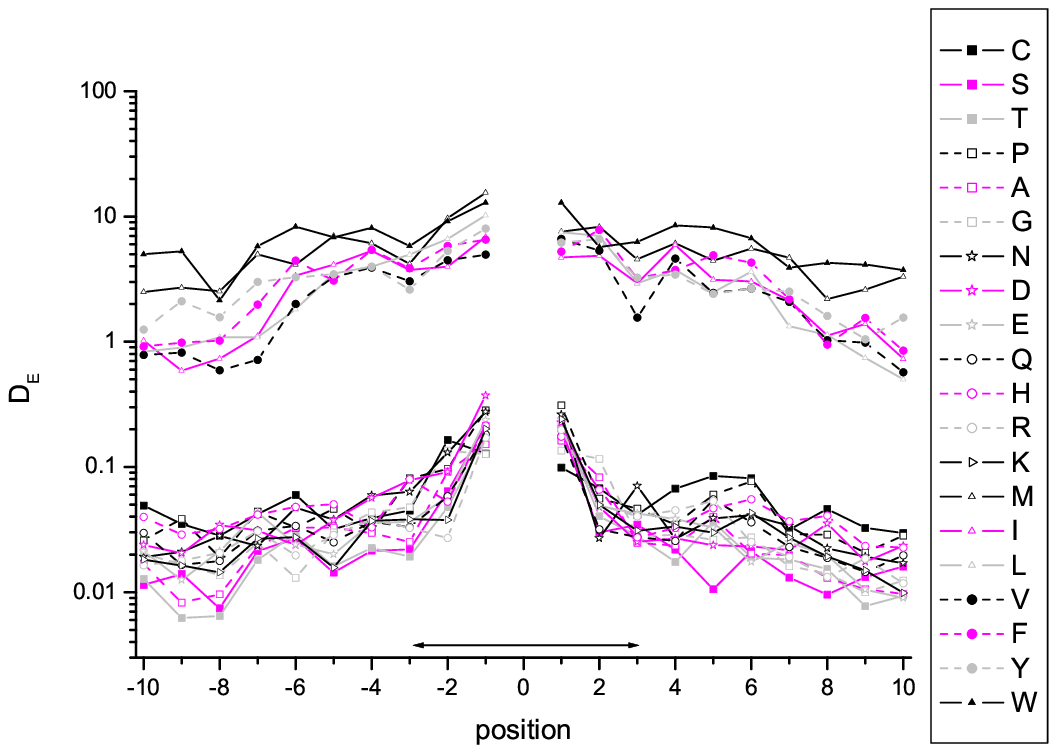}}
\caption{KL distances (doubled) of outer sites from their corresponding
noise background. Each curve is for an amino acid at the center labeled 0,
whose conformation is sheet. For clarity, the curves for M,I,L,V,F,Y and
W have been shifted up by multiplying an extra factor 100.} \label{fig3}
\end{figure}
\begin{figure}
\centerline{\epsfxsize=12cm \epsfbox{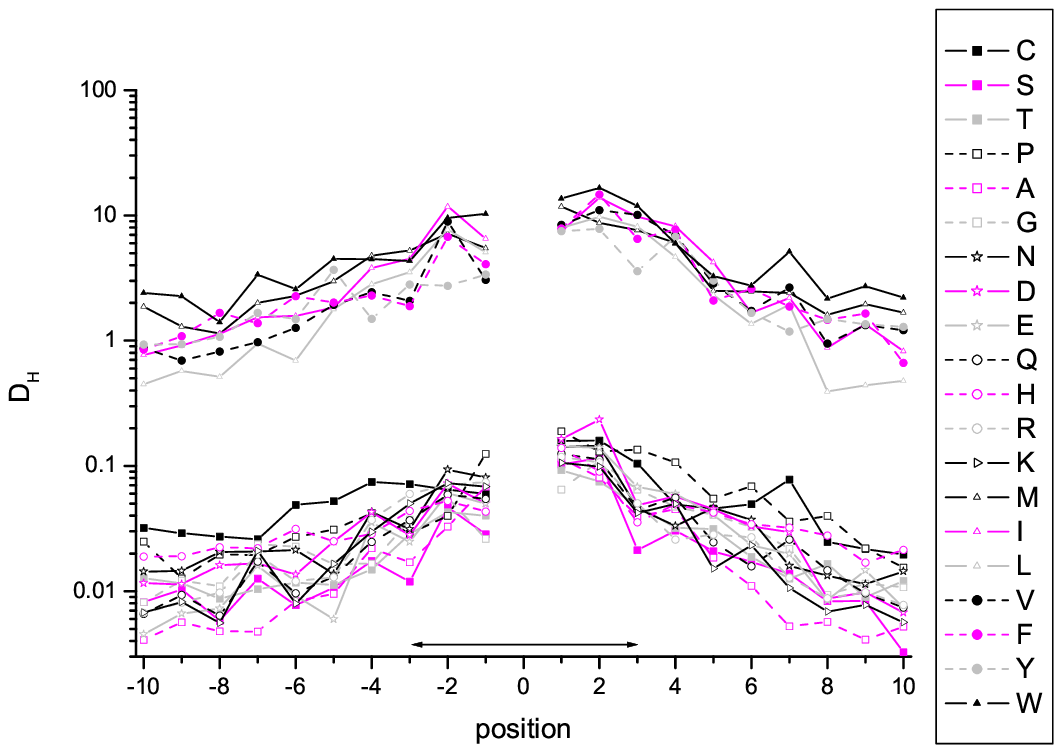}}
\caption{KL distances (doubled) of outer sites from their corresponding
noise background. Each curve is for an amino acid at the center labeled 0,
whose conformation is helix. For clarity, the curves for M,I,L,V,F,Y and
W have been shifted up by multiplying an extra factor 100.} \label{fig4}
\end{figure}

\begin{figure}
\centerline{\epsfxsize=12cm \epsfbox{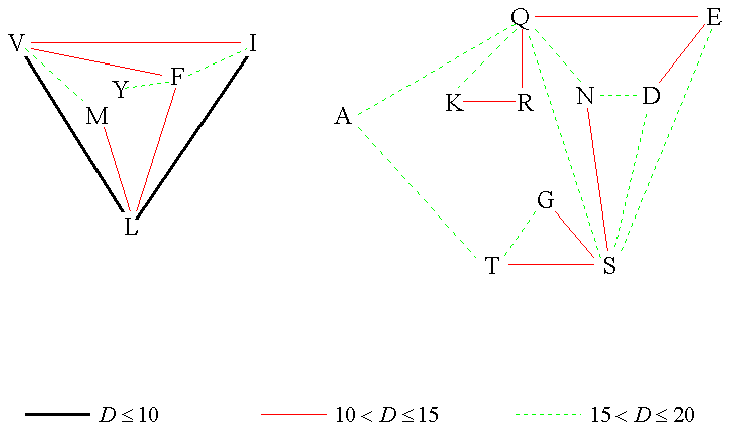}}
\caption{Connecting graph of amino acids in helix. Edges exist
only between vertices with a scaled distance not greater than 20.
Vertices without any connecting edges are not shown.}
\label{fig5}
\end{figure}

\begin{figure}
\centerline{\epsfxsize=12cm \epsfbox{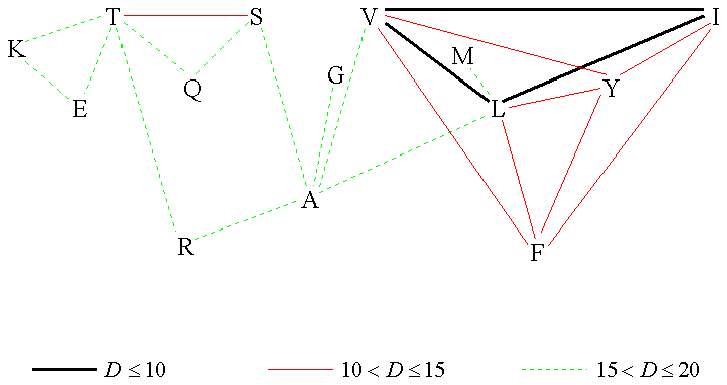}}
\caption{Connecting graph of amino acids in sheet. Edges exist
only between vertices with a scaled distance not greater than 20.
Vertices without any connecting edges are not shown.}
\label{fig6}
\end{figure}

\begin{figure}
\centerline{\epsfxsize=12cm \epsfbox{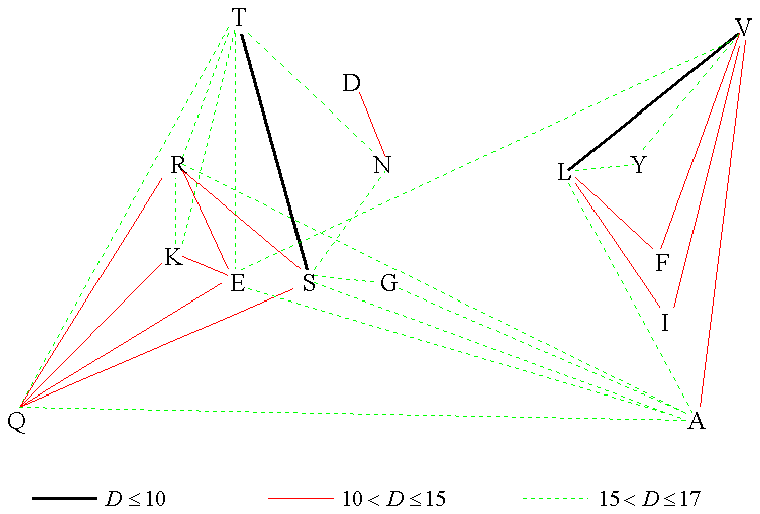}}
\caption{Connecting graph of amino acids in coil. Edges exist
only between vertices with a scaled distance not greater than 17.
Vertices without any connecting edges are not shown.}
\label{fig7}
\end{figure}

\begin{figure}
\centerline{\epsfxsize=12cm \epsfbox{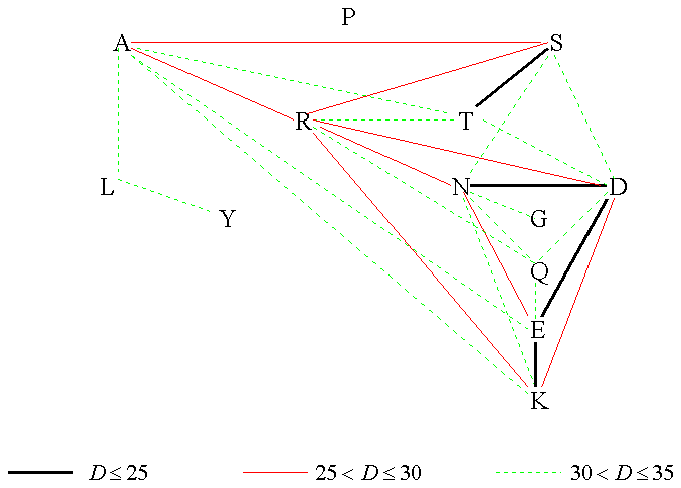}}
\caption{Connecting graph of amino acids in turn. Edges exist
only between vertices with a scaled distance not greater than 35.
Vertices without any connecting edges are not shown.}
\label{fig8}
\end{figure}

\end{document}